\documentclass{iaus}
\usepackage{graphicx}

\title[Secular evolution in galaxies] 
{Secular evolution in galaxies}

\author[F. Combes]   
{F. Combes}

\affiliation{LERMA, Observatoire de Paris, 61 Av. de l'Observatoire,
F-75014, Paris, France \break email: francoise.combes@obspm.fr}

\pubyear{2006}
\volume{235}  
\pagerange{119--126}
\date{24 August 2006  and in revised form 25 August 2006}
\jname{Galaxy Evolution across the Hubble Time}
\editors{F.Combes \& J. Palous, eds.}
\begin{document}

\maketitle

\begin{abstract}
New observations in favour of a significant role of secular evolution are reviewed: central star formation boosted in pseudo-bulge barred galaxies, relations between bulge and disk, evidence for rejuvenated bulges. Numerical simulations have shown that secular evolution can occur through a cycle of bar formation and destruction, in which the gas plays a major role. Since bars are weakened or destroyed in gaseous disks, the high frequency of bars observed today requires external cold gas accretion, to replenish the disk and allow a new bar formation.  The rate of gas accretion from external filaments is compatible with what is observed in cosmological simulations.
\end{abstract}

\firstsection 
\section{Role of secular evolution}

Secular evolution is one of the three scenarios for galaxy formation and evolution. In a first one, monolithic collapse, gas clouds in free fall form stars so quickly that the resulting system is ellipsoidal. Disks then form afterward, through gas accretion around these "bulges". In a second scenario, hierarchical formation, gas clouds have time to settle in rotating disks through dissipation, and form stars in thin spiral disks, which after several minor mergers, or a major merger, can also result in an ellipsoidal system. The secular evolution scenario also considers that spiral disks form first, then non-axisymmetric waves (bars and spirals) with their associated gravity torques and resonances, drive internal evolution, and in particular mass concentration towards the centre, and formation of bulges from the stellar disk. Spiral disks are continuously replenished through external gas accretion.

\subsection{Star formation and pseudobulges}

The bulges just formed out of disk stars in barred galaxies are likely to correspond to the observed 
 ``pseudobulges'', which have characteristics intermediate between a classical bulge (or Elliptical) and 
normal disks (Kormendy \& Kennicutt 2004). Their light distribution has a Sersic index n =1-2  (n=1 for disks and n=4 
for ellipticals), their flattening is also intermediate, with box-peanut shapes, and their kinematics 
reveal more rotation support than classical bulges.

Fisher (2006) has distinguished pseudobulges from classical ones in a sample of 50 galaxies from HST images. He used the PAH features in the 8 micron band as a tracer of star formation, from Spitzer observations. He founds that the central ($<$ 1.5kpc) star formation is more important  in galaxies with pseudo-bulges. Moreover, the effect is more spectacular when there exists a strong bar. Regan et al (2006) found similarly that pseudo-bulge galaxies of the SINGS sample are always associated with large central 8 micron emission, while there is no PAH concentration in classical bulge ones.

\subsection{Size of bars and stellar ages}

The observed relation between the bulge and disk characteristic radial scales
(MacArthur, Courteau \& Holtzman 2003) suggests a significant role for secular evolution. There is some scatter in the relation, but if bulges were entirely formed by minor mergers, no relation would be expected.
With respect to R$_{25}$, the bar radius ranges from 0.2 to 0.4 (Marinova \&  Jogee 2006), and bars are longer in early-types than late-types (Erwin 2005).

There is also a relation between metallicity and velocity dispersion in bulges, corresponding to the relation between metallicity and mass, but no relation age-metallicity as in elliptical galaxies. Bulges have a tendency to have smaller mean age, and a lower [$\alpha$/Fe] ratio, indicating a more continuous formation, with gas infall (Proctor et al 2000, Proctor \& Sansom 2002). A recent study by
Thomas \& Davies (2006) emphasizes that bulges are not in general very old, but rejuvenated systems. The influence of disks is particularly visible for late-types. Since the stars forming the pseudo-bulge are elevated from the disk by vertical resonances with the bar, the bulge stellar populations reflect those of the inner disk, which could be relatively old in early-type galaxies, or galaxies with little gas infall. This could explain the large range of abundances and ages.

Pseudo-bulges are observed in early-types as well: examples such as NGC 4593(SBb) and NGC 7690 (Sab) are given by Kormendy et al (2006). Starbursts are visible in these galaxies (without any companion), and it is suggested that star formation is episodic. Gas and dust are observed to accumulate in the nuclear ring of NGC 4593, which will form stars in the future.

\begin{figure}
\begin{center}
\includegraphics[width=0.85\textwidth]{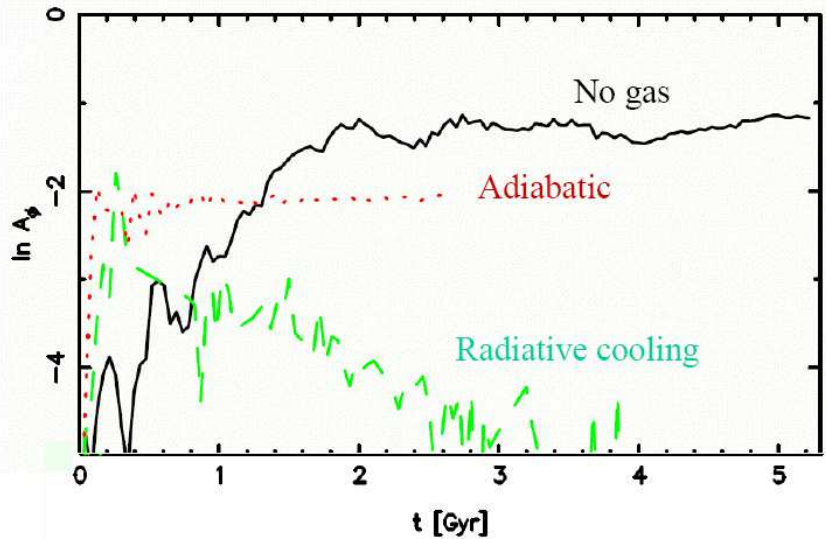}
\caption{Effect of gas in the evolution of bar strength, from Debattista et al (2006).
The top full (black) curve is a simulation without gas, the dotted (red) curve with adiabatic gas,
and the dash (green)  curve, with gas and radiative cooling.}
\label{deba}  
\end{center}
\end{figure}

\section{Bar destruction and reformation, role of gas}

It has been noticed for a long time, in many simulations of disk galaxies with a few percent of mass in gas, that the inflow of gas towards the center was weakening or destroying the bar (Friedli 1994, Berentzen et al 1998, Bournaud \& Combes 2002, Hozumi \& Hernquist 2005). However, it was believed that the weakening was due entirely to the central mass concentration (CMC) build in by the gas inflow. A CMC changes the inner mass distribution, destroys the orbital structure supporting the bar, and enhances the frequency of chaotic orbits (Hasan et al 1993). Recently it was emphasized that realistic CMC in disk galaxies are not able to destroy the bar (Shen \& Sellwood 2004, Athanassoula et al 2005), but the main destruction mechanism is instead the gas inflow itself. The gas is driven in by the bar torques, and reciprocally torques the bar. The angular momentum lost by the gas is taken up by the bar wave, which then weakens (Bournaud et al 2005).

The effect of gas depends on its physics, its dissipative character, and therefore on the assumed cooling: when isothermal, it is with more contrast that gas follows spiral arms in the barred potential, and the gravity torques are stronger. If it remains adiabatic, it has a behaviour more similar to the stellar component (Debattista et al 2006, figure \ref{deba}).

\begin{figure}
\begin{center}
 \includegraphics[width=0.7\textwidth]{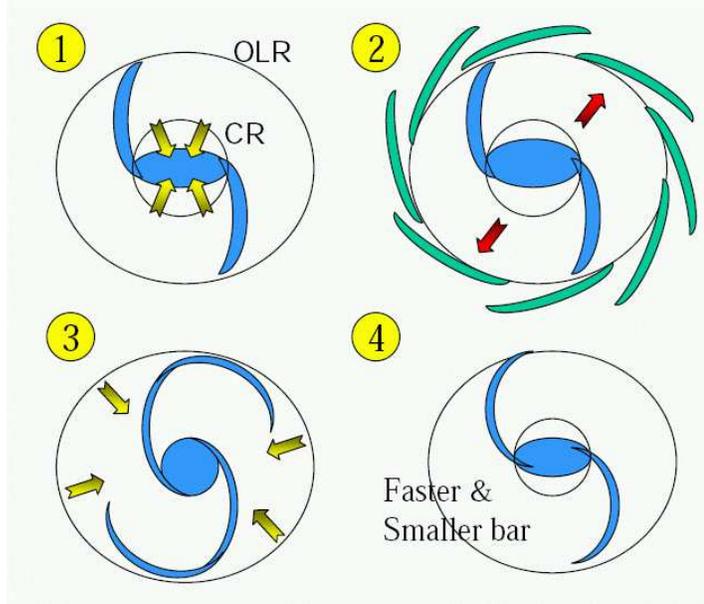}
  \caption{Schematic bar destruction/reformation cycle:
1-- A stellar bar forms in a cold gaseous disk,
and the bar gravity torques drive the gas inwards from corotation (CR)
at the outer end of the bar.
2-- As long as the bar is there, gas between CR and OLR (outer Lindblad
resonance) is driven outwards. The gas accreted from cosmic filaments is
stalled outside OLR.
3-- The bar is destroyed by the angular momentum accepted from the gas
inflow. Now bar torques are suppressed, and the external gas can inflow
and replenish the disk of this unbarred galaxy.
4-- A new bar instability develops in the new cold disk. Its pattern speed
is slightly higher, and the bar smaller with respect to the disk scalelength.}
\label{nuncius}
\end{center}  
\end{figure}

\section{Bar frequency and gas accretion}

Since bars waves are self-destroyed, it is possible to imagine a cycle of bar formation and destruction, such as described in figure \ref{nuncius}. A new bar can grow if the disk of cold gas reforms through external gas accretion. The observed frequency of bars can constrain the amount of gas accretion required.

Several authors have recently calculated the bar frequency from near-infrared samples, in particular the OSU NIR sample (Eskrige et al 2002). Whyte et al (2002) have computed the bar strength from the ellipticity of isophotes, and Block et al (2002) have computed the ratio $Q_b$ of the tangential to radial forces, from the potential derived from the red image. Both find a dearth of non-barred galaxies and a large abundance of strong bars. The paucity of weak bars is also found in different samples (Buta et al 2004, Marinova \& Jogee 2006): the majority of galaxies are barred (75\%) and most of them (80\%) are strongly barred. 

To maintain this high frequency of bars, continuous accretion at a large 
rate is required. The majority of the external gas cannot come from companions, since galaxy interactions heat the disks (Toth \& Ostriker 1992). What is required instead is a source of continuous cold gas accretion from the filaments in the near environment of galaxies. Cosmological accretion is compatible with doubling the mass in 10 Gyr (Semelin \& Combes 2005).

\section{Conclusion}

Secular evolution plays a fundamental role in mass assembly of spiral galaxies, through the intermediate of bars, pseudo-bulge formation, and external gas accretion. A gravity feedback mechanism self-regulate the evolution, and a cycle of bar formation/destruction can be accompanied by episodic starburst and AGN fueling.

Galaxy interactions and mergers play a major role in the formation of massive spheroids,
but the hierarchical scenario alone cannot explain the dominance in number of spiral galaxies, 
with young and thin disks. Essentially spiral galaxies are concerned by
secular evolution, and in particular those of intermediate types, with a small bulge.
Early-type galaxies can be re-transformed back in more late type objects through gas accretion, 
so can also be concerned by this scenario, according to their environment.


\end{document}